\documentclass[twocolumn,aps,showpacs,superscriptaddress,showkeys,floatfix]{revtex4} 
\usepackage{graphicx}
\begin{document}
\title{Dynamics of Emitting Electrons in Strong Electromagnetic Fields} 
\author{Igor V. Sokolov}
\email{igorsok@umich.edu}
\affiliation{Space Physics Research Laboratory, University of Michigan, Ann
Arbor, MI 48109 }
\author{Natalia M. Naumova}
\affiliation{Laboratoire d'Optique Appliqu\'{e}e, UMR 7639 ENSTA, Ecole
Polytechnique, CNRS, 91761 Palaiseau, France}
\author{John A. Nees}
\affiliation{Center for Ultrafast Optical Science and FOCUS Center, University
of Michigan, Ann Arbor, MI 48109}
\author{G\'{e}rard A. Mourou}
\affiliation{Laboratoire d'Optique Appliqu\'{e}e, UMR 7639 ENSTA, Ecole
Polytechnique, CNRS, 91761 Palaiseau, France}
\author{Victor P. Yanovsky}
\affiliation{Center for Ultrafast Optical Science and FOCUS Center, University
of Michigan, Ann Arbor, MI 48109}
\date{\today}
\begin{abstract}
We derive a modified non-perturbative Lorentz-Abraham-Dirac equation. It 
satisfies the proper conservation laws, particularly, it conserves 
the generalized momentum, the latter property eliminates the 
symmetry-breaking runaway solution. The equation allows a consistent 
calculation of the electron current, the radiation effect on the electron 
momentum, and the radiation itself, for a single electron or plasma
electrons in strong electromagnetic fields. 
The equation is applied to a simulation of a strong laser pulse 
interaction with a plasma target. Some analytical solutions are also
provided. 
\end{abstract}
\maketitle
\section{Introduction}
Lasers now allow us to reach intensities within the 
focal spot of $W> 10^{22}\ {\rm W/cm}^2$ 
\cite{1022}. Electron motion in fields where $W\gg 10^{18}\ {\rm W/cm^2}$, 
for a typical laser wavelength, $\lambda\sim1\mu$m, is 
ultra-relativistic:
\begin{equation}\label{A0}
a=\frac{|e|A}{mc^2}\gg1,
\end{equation}
where $e$ is the electron charge, $m$ is its mass, 
$c$ is the speed of light and $A$ is the 
vector-potential amplitude.

An accelerated electron in a strong laser field emits 
high-frequency radiation \cite{sarachik}. 
Its back-reaction on the electron motion can not be 
neglected, if 
in the 
frame where the electron is initially at rest,
the energy radiated during the interaction time is 
comparable with  
$mc^2$: $\frac{\sigma_T}{4\pi}\int {\bf E}^2cdt\ge mc^2$, where 
$\sigma_T=\frac{8\pi e^4}{3m^2c^4}$ and ${\bf E}$ is the electric field.  
In the 
course of a Lorentz transformation, this integral transforms 
proportionally to a wave frequency, $\omega_0$. Indeed, 
${\bf E}\sim \omega_0 {\bf A}$. The transversal 
vector potential, ${\bf A}$, as well as the differential of the wave 
phase, $\omega_0 dt$, are Lorentz invariant. Since 
$\omega_0{\cal E}-c^2({\bf k}_0\cdot{\bf p})=c^2(k_0\cdot p)$ is 
invariant, as is the 4-dot-product of the particle 
momentum, $p^i=(\frac{\cal E}c,{\bf p})$, 
by the wave number, $k^i_0=(\frac{\omega_0}c,
{\bf k}_0)$, the (sufficient)
condition for the radiation reaction significance is as follows: 
\begin{equation}\label{main}
\frac{\sigma_T}{mc^2}
\int {\bf E}^2\frac{cdt}{4\pi}=
\frac{\int{Wdt}}{1.2\ {\rm kJ}/(\mu{\rm m})^2}\ge \frac{mc^2}{{\cal E}-cp_x}.
\end{equation}
Here ${\cal E}$ and ${\bf p}$ are the particle energy and momentum and the 
wave is assumed to propagate along the $x$-axis. 
A high value of the integral in (\ref{main}) may be 
reached, in principle, at the cost of higher intensity only, 
$W\sim10^{25}\ {\rm W/cm}^2$. In the course of the 
ELI  project (see \cite{ELI}) a laser is expected to reach 
focusable pulse energy of 1.5 kJ at $\lambda\approx 0.8\mu$m, so the 
radiation effects will be dominant: 
$\int{Wdt}\approx 2.1\ {\rm kJ}/(\mu{\rm m})^2
\ge 1.2\ {\rm kJ}/(\mu{\rm m})^2$. Another opportunity may be realized while a 
strong laser pulse interacts with  energetic electrons, which 
move {\it oppositely} to the direction of the pulse propagation. In this case 
${\cal E}-p_x c\approx 2{\cal E}\gg mc^2$,
facilitating the fulfillment of Eq.(\ref{main}).
In such a geometry, powerful X-ray radiation is generated in the direction 
of the electron momentum \cite{spres}. 
Ineq.(\ref{main}) determines the regime, in which the
energy is {\it efficiently} converted to X-ray or $\gamma$ bursts.

In the course of a strong laser pulse interacting 
with a dense plasma the 
counter-propagating electrons may be accelerated in a backward direction by a 
charge separtion field (see, e.g., \cite{nat04b}). At moderate intensities 
the generation of short pulses of  higher-frequency 
radiation \cite{nat04a} may be interpreted as the reflection of 
the laser pulse from these bunches as from a reflecting medium:  
the frequency of the reflected wave, $\omega^{(r)}$, is upshifted: 
$\omega^{(r)}\sim(\gamma^{(m)})^2\omega_0$, if the reflector moves with a 
Lorentz-factor of $\gamma^{(m)}\gg 1$ towards an incident wave.
Here we consider such high laser intensities that
emission frequencies are upshifted to the hard X-ray and $\gamma$ 
range. At realistic electron density,  
$N_e\le10^{24}{\rm cm}^{-3}$,  the averaged field approximation of the 
reflecting medium is not applicable for emitted photon energies exceeding 
10 keV, because $N_e(c/\omega^{(r)})^3\ll 1$. The emission from plasma in this 
case is taken as an integral of the radiation intensity from 
separate electrons, rather than as the field of a coherent electric current 
of a $\gamma$ range frequency in plasma. Hence, even in a dense plasma the
emission from 
separate electrons is essential to the analysis.

While the physical processes involving 
significant radiation back-reaction 
are of growing importance, the Lorentz-Abraham-Dirac (LAD) model 
\cite{dir38} which is to 
account for this effect, is not free of difficulties, such as  
`runaway' solutions {\it etc.} 
\cite{dir38,ll,pan,jac,el,kl,poi99,panofsky}.
Some flaws of the original Dirac version 
(see Eq.(\ref{LAD}) below), are eliminated in the 
approximate equation, as derived in the book by Landau and Lifshits \cite{ll} 
as Eq.(76.3), see also the non-relativistic variant 
in \S 75. 
A slightly different 
approximation was found by 
Eliezer \cite{el}, 
and most later versions (see, e.g., 
\cite{kl,poi99,panofsky}) 
are reducible to those listed above.

New problems arise while considering the 
transition to even higher field intensities, at which 
Quantum ElectroDynamical 
(QED) effects come into a power. Delegating the 
discussion of 
QED strong fields to a forthcoming publication, we still 
have to briefly discuss here a more general issue of the LAD model conformity 
with the QED principles. Particularly, the nature of generalized electron 
momentum (which is substituted for the $i\hbar\nabla$ operator in the framework 
of QED) is highly argueable in the LAD model, but this point has hardly been
discussed so far. 
Also, there is a controversy between the treatment of 
{\it the radiation} in  QED as a {\it random process} of separate photon
emission with some probability, and the description of 
{\it the back-reaction } of the radiation in the LAD model with 
very {\it smooth functions} of time which are allowed to be differentiated
many times.   

In Section II we describe
how we derive the modified LAD equation for 
electrons and account for radiation from electrons and the 
electron current in a plasma, in a self-consistent manner 
and in the way which does not contradict 
QED fundamentals. 
As an application, the basic elements of a
particle-in-cell (PIC) 
numerical scheme are discussed, 
which 
extends the simulation of laser-plasma interactions to 
higher intensities $W\ge 10^{22}\ {\rm W/cm}^2$.
In Section III we analytically solve 
for the electron motion in 
a 1D wave 
field in vacuum.
Results of 3D PIC simulations of laser-plasma 
interaction 
at intensities $W\sim10^{22}\ {\rm W/cm}^2$ 
are discussed in the
concluding section. 
\section{Modified LAD model}
Here we assume an electron  moving in an external electromagnetic 
field and emitting high-frequency radiation. In the case of a plasma
electron, the external field is the averaged field as present in the 
Maxwell-Vlasov equations. We derive the electron momentum 
equation 
and discuss its 
(minor) differences from the Dirac theory \cite{dir38}.

{\bf Illustration: external field of the 1D wave.} 
We start from an example of the field of a 1D harmonic wave with the wave 4-vector, $k^i_0$.
Recall that the energy-momentum
exchange of the charged particle with the classical field is governed by the 
Lorentz force, while the effect of the emitted/absorbed photons should be interpreted in 
terms of the photon 4-momentum. The case of the 1D wave allows
both treatments. 

In the course of photon emission with the 4-vector, $k^i_1$, 4-momentum is conserved: 
$p^i_1=p_0^i+n\hbar k^i_0-\hbar k^i_1$, where $p^i_{0,1}$ stand for 4-momenta 
of
the electron 
before and after the emission. In the classical limit of small recoil, 
the increment in the electron 
momentum, 
$\delta p^i=p_1^i-p_0^i$, should be small. Therefore, the condition 
$(p\cdot p)=({\cal E}/c)^2-{\bf p}^2=m^2c^2$ requires that $(\delta p\cdot p)=0$ 
and $\delta p^i=\hbar k_0^i\frac{(k_1\cdot p_0)}{(k_0\cdot p_{0})}-\hbar k_{1}^i$. 
The
second term is the 4-momentum transferred to the emission and
the first term is 
the gain in 4-momentum the electron obtained from 
the field. The latter for the classical external field 
reduces to 
the effect of the field tensor, $F^{ik}$,  
on the yet 
unknown current, $e\delta (dx^i/d\tau)$:
\begin{equation}\label{eq:newcurrent}
\int_{\Delta \tau}{\frac{e}cF^{ik}
\delta (\frac{dx_k}{d\tau})d\tau}=\hbar k_0^i
\frac{(k_1\cdot p_0)}{(k_0\cdot p_{0})}.
\end{equation}
Here $\tau$ is the  time in the `Momentarily Comoving' Lorentz Frame (MCLF), 
such that the spatial components of $p_0^i$ vanish. In  
strong fields as in 
Eq.(\ref{A0}), 
emission characteristics are local functions of the wave  
field (see \cite{lp}, \S 90). Therefore, the integration in 
Eq.(\ref{eq:newcurrent}) reduces to a multiplication by $\Delta \tau$. 
Expressing $F^{ik}=(\partial A^k/\partial x_i)- (\partial A^i/\partial x_k)$, 
in terms of the 4-vector-potential and using the entities,    
$(k_0\cdot A)=0$, 
$(k_0\cdot k_0)=\omega_0^2/c^2-{\bf k}^2_0=0$ one can represent 
$\frac{k_0^i}{(k_0\cdot p_{0})}=
\frac{F^{ik}F_{kl}p^l_{0}}
{p^i_{0}F_i^{\,k}F_{kl}p^l_{0}}$  and solve 
Eq.(\ref{eq:newcurrent}):
\begin{equation}\label{eq:se}
\delta (\frac{dx^i}{d\tau})=
-\frac{(p_0\cdot \hbar k_1) f^i_{L0}}{m(f_{L0}\cdot f_{L0})
\Delta\tau},\,\,\,\,f^i_{L0}=\frac{eF^{i}_lp^l_{0}}{mc}.
\end{equation}
In the MCLF the current has only spatial components and may be expressed in 
terms of the Lorentz transformed electric field, ${\bf E}_{\rm MCLF}$:
$\delta \left(\frac{d{\bf x}}{d\tau}\right)=
\frac{\hbar\omega_1{\bf E}_{\rm MCLF}}{eE^2_{\rm MCLF}
{\Delta\tau}}$, $e^2E^2_{\rm MCLF}=-(f_{L0}\cdot f_{L0})$. 
Hence, the emission 
is accompanied by the displacement of the electron  along 
$e{\bf E}_{\rm MCLF}$.  

Now we average  Eqs.(\ref{eq:se}) over the emitted photon parameters. The
averaging (taking a mathematical expectation) is done as a weighted 
integration over $d\omega_1$ with the differential probability of emission 
per unit of time, $dW/d\tau d\omega_1$, the result being multiplied by $\Delta \tau$, 
to account for the time integration. In the MCLF, averaging of 
$\hbar\omega_1$ gives $I\Delta\tau$, where
$I=\int_{\omega_{\rm min}}^\infty{\hbar 
\omega_1\frac{dW}{d\tau d\omega_1}}d\omega_1$
is the total emission intensity.  Below we use its ratio to the dipole emission 
intensity, $I_E=\tau_0e^2E_{\rm MCLF}^2/m$, 
$\tau_0=2e^2/(3mc^3)\sim 6.2\cdot10^{-24}$ s. In an arbitrary frame of 
reference,  averaged $\hbar k_1^i$ is the 4-momentum of emitted radiation, 
$(\frac{d p^i}{d\tau})_{\rm rad}\Delta\tau$, 
expressed in terms of $I$: 
$(\frac{d p^i}{d\tau})_{\rm rad}=\frac{p_0^i}{mc^2}I$ (see \cite{ll}, \S 73).
Analogously, averaging $(p_0\cdot\hbar k_1)$ gives $mI\Delta\tau$, so 
that the averaged Eq.(\ref{eq:se}) reads: 
$(
\frac {dx^i}{d\tau})_{\rm rad}=
\tau_0\frac{ I}{I_E}\frac{ f^i_{L0}}{m}$ and 
the momentum
equation for electron becomes:
\begin{equation}\label{eq:dpdtau}
\frac{dp^i}{d\tau}=
eF^{ik}\frac{p_k}{mc}-\frac{p^iI}{mc^2}+\tau_0e^2\frac{I}{I_E}
\frac{F^{ik}F_{kl}p^l}
{(mc)^2},
\end{equation}
where the  terms 
on the right hand side are: the Lorentz force, $f_{L0}^i$,  
the 4-momentum of the emitted radiation, $-(\frac{d p^i}{d\tau})_{\rm rad}$, 
and the external field effect, $F^{i}_{\,k}
J^k_{\rm rad}/c$, 
on the current, $J^k_{\rm rad}=e(\frac {dx_k}{d\tau})_{\rm rad}$.  
Multiplying Eq.(\ref{eq:dpdtau}) by $p_i$ we see that $d(p\cdot p)/d\tau=0$, maintaining the entity, 
$(p\cdot p)=m^2c^2$.

{\bf General case of an arbitrary external field.} Eq.(\ref{eq:dpdtau}) is not 
specific to the 1D wave case and can be derived for an arbitrary external
electromagnetic field. Seeking the last term in the form of 
$F^{i}_{\,k}J^k_{\rm rad}$, which is mandatory for the 4-momentum exchange with the 
classical external field, and requiring the conservation of $(p\cdot p)$ we 
obtain Eq.(\ref{eq:dpdtau}) directly and with no extra assumption.

{\bf Electron current.}  Now we re-write Eq.(\ref{eq:dpdtau}) in terms of
the {\it total electron current}, 
$e\frac{dx^i}{dt}=
e\frac{p^i}m+e\left(\frac {dx^i}{d\tau}\right)_{rad}$:
\begin{equation}\label{eq:pwithj}
\frac{dp^i}{d\tau}=\frac{e}{c}F^{ik}\frac{dx_k}{d\tau}-\frac{Ip^i}{mc^2},
\end{equation}
\begin{equation}\label{eq:dxdtau}
\frac{dx^i}{d\tau}=\frac{p^i}{m}+\tau_0\frac{I}{I_E}
\frac{eF^{ik}p_{k}}
{m^2c}.
\end{equation}
Integrating by volume the equation for the 
energy-momentum tensor for the external field, 
$\partial{T^{ik}_{\rm ext}}/{\partial x^k}=-\frac1cF^{ik}j_k$, and 
representing the volume integral, $\int{j^idV}$, of the point-wise current density, $j^i=ec\int{\frac{dx^i}{d\tau}\delta^4( r^k-x^k(\tau))d\tau}$,  
($r^k$ being the coordinate 4-vector in an arbitrary Lorentz frame) 
in terms of $e\frac{dx^i}{d\tau}$, we find:
$
\frac{d}{d\tau}{\int{ T^{i0}_{\rm ext}dV}}=
-\frac{e}{c}F^{ik}\frac{dx_k}{d\tau}
$ 
(cf \cite{ll}, \S 33). Hence, 
Eqs.(\ref{eq:pwithj}-\ref{eq:dxdtau}) 
conserve the total 
energy-momentum: 
$\frac{dp^i}{d\tau}+\frac{d}{d\tau}{\int{T^{i0}_{\rm ext} dV}}+\frac{Ip^i}{mc^2}=0$. 

The generalized momentum may
also be conserved. However, for 
a radiating electron this conservation takes place, if not only the external field is 
constant along some direction: $({\bf n}\cdot\nabla) A^i=0$, but also the 
projection of the emitted momentum, $(\frac{d p^i}{d\tau})_{\rm rad}$, onto 
${\bf n}$ vanishes:  $I({\bf p}\cdot {\bf n})=0$. If the latter 
condition is not fulfilled and $I({\bf p}\cdot {\bf n})\ne 0$, then 
the change 
in the generalized momentum, ${\cal P}^i=p^i +eA^i/c$,  is as follows:
\begin{equation}\label{eq:gen}
\left({\bf n}\cdot \frac{d{\bf {\cal P}}}{d\tau}\right)=-\left({\bf n}\cdot
{\bf p}\right)\frac{I}{mc^2}.
\end{equation} 
Eq.(\ref{eq:gen}) also follows from the quantum relationship, 
${{\bf n}\cdot\delta{\bf {\cal P}}}=-\hbar {{\bf n}\cdot{\bf k}_1}$, 
(a conserved generalized momentum corresponds to the constant gradient of the 
electron wave function phase along ${\bf n}$ -  see \cite{lp}). 

Discussing possible choices of $I$,
we note that the ratio $I/I_E$ should be bounded at 
$I_E\rightarrow 0$. Although to take $I=I_E$ 
is physically reasonable, there are 
other interesting options. Particularly, $I$ can be a random function with its 
average equal to $I_E$ (to trace the quantum theory limit or to include emission with 
large photon energy). One can apply $I$ expressed in terms of the modified emission 
probability, to treat the processes (like a gyrosynchrotron emission, see \S90 in \cite{lp}) in very 
strong fields, such that the QED effects are not negligible. After all, $I$  
can differ  from $I_E$ by a choice of $\omega_{\min}\ne 0$. The latter 
approach allows us to separate, if desired, the high-frequency emission from a 
lower-frequency averaged external field, in simulating laser-plasma 
interactions.

Eqs.(\ref{eq:pwithj}-\ref{eq:dxdtau}) 
and their properties result from the assertion that the electron while
emitting moves not strictly along the direction of its momentum.  
Particularly, in the MCLF the electron while emitting is not at rest 
and displaces along $e{\bf E}_{\rm MCLF}$. In the MCLF 
the external electric field produces a work at a moving charge,  
which {\it entirely} balances the emitted energy: 
$e(\frac{d{\bf x}}{d\tau}\cdot {\bf E}_{\rm MCLF})=I$. 

The model applicability is limited by the requirement for the current 
$(dx^i/d\tau)_{\rm rad}$ to be essentially non-relativistic, which is 
fulfilled as long as $\tau_0I^2/I_E\ll mc^2$. To neglect QED effects,
the field should be weak:
$eE_{\rm MCLF}\ll mc^2(mc/\hbar)$, and  $\tau_0I_E/mc^2\ll (e^2/\hbar c)^2\ll1$.

{\bf To compare with the radiation force model} we
use MCLF and approximate within a short time interval ${\bf p}=0$ 
and put $I=I_E$ in spatial components of 
Eqs.(\ref{eq:pwithj},\ref{eq:dxdtau}): 
$d{\bf p}/d\tau\approx e{\bf E}+\tau_0 e^2[{\bf E}\times{\bf B}]/(mc)$,
$d{\bf x}/d\tau\approx ({\bf p}+\tau_0 e{\bf E})/m$. Formally, the latter is 
equivalent to the Newton equation with the approximate 
radiation force as described in \cite{ll,panofsky,poi99}:
$md^2{\bf x}/d\tau^2=e{\bf E}+\tau_0(ed{\bf E}/d\tau+
e^2[{\bf E}\times{\bf B}]/(mc))$. 

Now we compare Eqs.(\ref{eq:pwithj}-\ref{eq:dxdtau}) with the LAD equation \cite{dir38}:
\begin{equation}\label{LAD}
\frac{d^2x^i}{d\tau^2}=\frac{e F^{ik}}{mc}\frac{dx_k}{d\tau}+\tau_0
\frac{d^3x^i}{d\tau^3}+\frac{\tau_0}{c^2}\frac{dx^i}{d\tau}\left(\frac{d^2x}{d\tau^2}
\cdot\frac{d^2x}{d\tau^2}\right),
\end{equation}
(cf. \cite{ll}, Eqs.(76.1-2)). 
Re-write Eq.(\ref{LAD}) introducing $I=-m\tau_0(d^2x_k/d\tau^2)(d^2x^k/d\tau^2)$, 
as in \cite{ll}, Eq.(73.4):
$$
\frac{dp^i_{\rm D}}{d\tau}
=\frac {e F^{ik}}c\frac{dx_k}{d\tau}-\frac{dx^i}{d\tau}\frac{I}{c^2},
\,\,\,\frac{ p^i_{\rm D}}m=
\frac{dx^i}{d\tau}-\tau_0\frac{d^2x^i}{d\tau^2}.
$$
Comparing this with Eqs.(\ref{eq:pwithj}-\ref{eq:dxdtau}) we find that
both our model and the Dirac theory, as well as the modified version of 
Eq.(\ref{LAD}) as described in \cite{ll,panofsky,poi99} 
(which approximates 
$\frac{d^2x^i}{d\tau^2}\approx\frac{e F^{ik}}{mc}\frac{dx_k}{d\tau}$ 
in the right hand side of Eq.(\ref{LAD})) differ from each other with small 
terms $\sim\tau_0^2$.

The key distinction, however, is the choice of the electron momentum. 
An interesting survey \cite{poi99} shows that this choice in 
the Dirac theory is ambiguous. It is problematic too:  
for $p^i=mdx_i/d\tau$ Eq.(\ref{LAD}) conserves $(p\cdot p)=m^2c^2$, but 
the generalized momentum is not conserved, however symmetric the 
external field and the radiation 
may be. Particularly, Eq.(\ref{LAD}) 
allows  {\it this} electron momentum to change in the absence of 
the external field (the runaway solution, see \cite{ll,panofsky,poi99}),  while 
the conservation of the generalized momentum in the MCLF would enforce 
${\bf {\cal P}}={\bf p}=0$, as long as 
$({\bf n}\cdot \nabla)A^i=0$ for any ${\bf n}$ and $I=0$.  
With a different choice of the momentum 
(say, $p^i_{\rm D}$ as introduced above) the generalized momentum may 
conserve, but 
not $(p\cdot p)$.
So, the distinction of our approach from the Dirac model lies in: 
(1) 
the incorporation of $\sim(\frac{dx^i}{d\tau})_{\rm rad}$ into the relationship between
the velocity, $\frac{dx^i}{d\tau}$, and momentum, $p^i$ instead of the
``self-force'' $\sim\frac{d}{d\tau}(\frac{dx^i}{d\tau})_{\rm rad}$ into the
force equation
for  
$dp^i/d\tau$ and (2) the use of a different relativistic formulation, 
providing a different set of exact conservation laws.

{\bf Application to the particle-in-cell scheme.}
3-vector formulation of Eqs.(\ref{eq:pwithj}-\ref{eq:dxdtau}) is as simple as:
\begin{equation}\label{eq:3plus1}
\frac{d{\bf p}}{dt}={\bf f}_L+\frac{e}c[\delta{\bf u}\times{\bf B}]-
\frac{{\bf u}\gamma^2}{c^2}(\delta{\bf u}\cdot {\bf f}_L),\,\,\
\frac{d{\bf x}}{dt}={\bf u}+\delta{\bf u},
\end{equation}
where: $I=I_E$, 
$
{\bf u}=\frac{\bf p}{\sqrt{m^2+{\bf p}^2/c^2}}$, ${\bf f}_L=e{\bf E}+\frac{e}c[{\bf u}\times{\bf B}]$, and
\begin{equation}
\delta{\bf u}=\frac{\tau_0}m\frac{{\bf f}_L-{\bf u}({\bf u}\cdot {\bf f}_L)/c^2}{1+\tau_0({\bf u}\cdot {\bf f}_L)/(mc^2)}.
\end{equation}
These equations may be applied to plasma electrons in order to simulate  
laser-plasma interactions at high laser field intensity. More precisely, 
the equations should be solved for `particles' consisting of a large number of 
electrons, radiating independently and incoherently. Their contribution into 
an averaged electron current is $e({\bf u}+\delta{\bf u})$. The spectrum of radiation 
from relativistic electrons is calculated assuming that an angular
distribution 
is peaked in the direction of the electron momentum and can be approximated with the $\delta$-function
and the frequency spectrum, 
$F(r)=\frac{3^{5/2}}{8\pi}r\int_r^\infty{K_{5/3}(r^\prime)dr^\prime}$, $r=\omega_1/\omega_c$, $\omega_c=\frac32\omega_r\gamma^3$,
is momentarily close to that from 
circular motion with 
a rotation frequency, 
$\omega_r=|{\bf p}\times{\bf f}_L|/{\bf p}^2$:
\begin{equation}
\frac{dI}{d{\bf \Omega} d\omega_1}\Delta t=\delta\left({\bf \Omega}-\frac{\bf p}{|{\bf p}|}\right)\frac{\gamma^2(\delta{\bf u}\cdot {\bf f}_L)}{\omega_c}F\left(\frac{\omega_1}{\omega_c}\right)\Delta t.
\end{equation}

The integral of the spectral function is normalized by unity,
$\int{F(x)dx}=1$. 
The effect of the radiation on the electron motion, 
$\int{ 
\frac{dI}{d{\bf \Omega} d\omega_1}
d{\bf \Omega} d\omega_1}=\gamma^2(\delta{\bf u}\cdot {\bf f}_L)$, is entirely included into Eq.(\ref{eq:3plus1}).

\section{Electron in the 1D wave}
In the case of the 1D wave external field, 
the electron motion can be solved analytically. 
With the external field being a function of 
$\xi=(k_0\cdot x)$, the relation between $\xi$ and $\tau$ is given by
a product of Eq.(\ref{eq:dxdtau}) by $k_0$: $d\xi/d\tau=(k_0\cdot p)/m$. 
Multiplying Eq.(\ref{eq:pwithj}) by $k_0$, expressing the derivative over 
$\tau$ in terms of that over $\xi$ and assuming 
$I=I_E=\tau_0(k_0\cdot p)^2c^2|d{\bf a}/d\xi|^2/m$, we obtain: 
$d(k_0\cdot p)/d\xi=-(k_0\cdot p)^2\tau|d{\bf a}/d\xi|^2/m$, and (cf. to Eq.(2)): 
\begin{equation}\label{eq:kcdotp}
\frac1{(k_0\cdot p)}=\left(\frac1{(k_0\cdot p)}\right)_{\xi=0}+
\frac{\tau_0}m\int_0^\xi{\left|\frac{d{\bf a}(\xi_1)}{d\xi_1}\right|^2d\xi_1}.
\end{equation}  
The transverse momentum, ${\bf p}_\perp$, is solved 
from Eq.(\ref{eq:gen}):
\begin{equation}
\frac{{\bf p}_\perp+\frac{e{\bf A}}c}{(k_0\cdot p)}=
\left(\frac{{\bf p}_\perp+
\frac{e{\bf A}}c}{(k_0\cdot p)}\right)_{\xi=0}+
c\tau_0\int_0^\xi{{\bf a}\left|\frac{d{\bf a}}{d\xi_1}\right|^2d\xi_1}.
\end{equation}
To compare with the 
Dirac solution for a short pulse \cite{dir38}, consider a single-period symmetric wave: ${\bf a}={\bf a}_0\sin(\xi)$, $0<\xi<2\pi$. 
In the frame of reference in which the electron was at
rest prior to the interaction the transverse components of the electron 
momentum vanish after the pulse: not only they do not turn to infinity, as 
they would in the Dirac runaway solution, but the conservation of the generalizied 
momentum in conjunction with the pulse symmetry ($\int_0^{2\pi}{{\bf a}\left|\frac{d{\bf a}}{d\xi_1}\right|^2d\xi_1}=0$) entirely eliminates ${\bf p}_\perp$ at $\xi>2\pi$. For a pulse 
of moderate intensity, such that the left hand side of Eq.(2) is much less 
than unity, the electron gains a small momentum,
$p_x=\sigma_T\int{{\bf E}^2dt}/(4\pi)$, in the direction of the pulse 
propagation, this momentum is absorbed from the pulse. The energy, $cp_x$, 
is absorbed from the pulse and almost equal energy is emitted.
\begin{figure}
\includegraphics[scale=0.55]{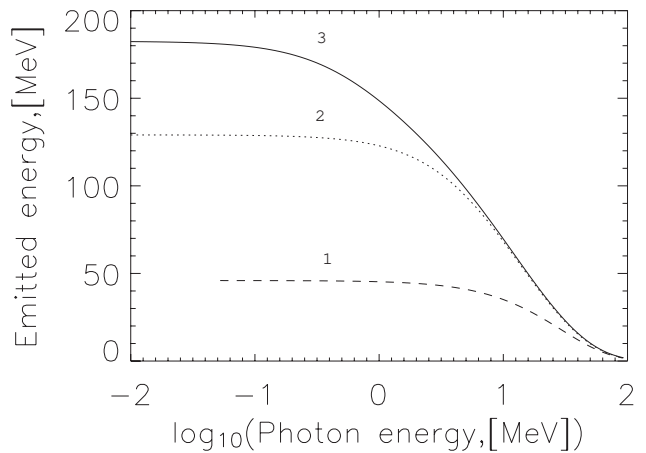}
\caption{Energy of backward emitted photons 
with $\hbar\omega^\prime>E$, as a function of $E$, where a circularly polarized wave of the amplitude of $a_0=50$, 
interacts with a 
counter-propagating electron of the energy of ${\cal E}=180$MeV, for different pulse durations:
2T (curve 1), 14T (2) and 81T  (3), $T\approx2.7$ fs.
}
\label{fig_1}
\end{figure} 
For strong pulses satisfying Ineq.(2), we present an 
integral spectrum of emission in Fig.1. 
We see that for a longer pulse duration the spectrum is softened. This is a result of
the radiation reaction: without it the spectrum would have the same
shape as that of 
the shorter 5-fs pulse and would only increase proportionally to the pulse
duration.

\section{Simulations and discussion}   
To demonstrate more realistically the role of the radiation back-reaction in 
the laser-plasma interaction we perform a 3D PIC simulation for 
a 10-cycle linearly polarized laser pulse 
having a step-like profile along the pulse 
direction 
including
2-$\lambda$ rising and falling edges, and a Gaussian profile in the transverse 
direction, focal diameter $5\lambda$, and  amplitude $a_0=70$. 
The laser pulse is incident normally on a
plasma layer of  10-$\lambda$ length
and density $n_0=3n_{cr}$
where $n_{cr}=mc^2\pi/(\lambda e)^2$ is the critical density. 
The simulation is performed in the box $20\lambda\times
20\lambda\times 20\lambda$ with spatial resolution $\lambda/20$ and
8 electrons per cell, 
requiring in all
$6.4\times 10^7$ grid cells and
$2.6\times 10^8$ particles. The plasma layer is located after
a 5-$\lambda$ vacuum layer. 
Here ions are immobile and the time step is 
$\Delta t=\lambda/(40c)$.
\begin{figure}
\includegraphics[scale=0.8]{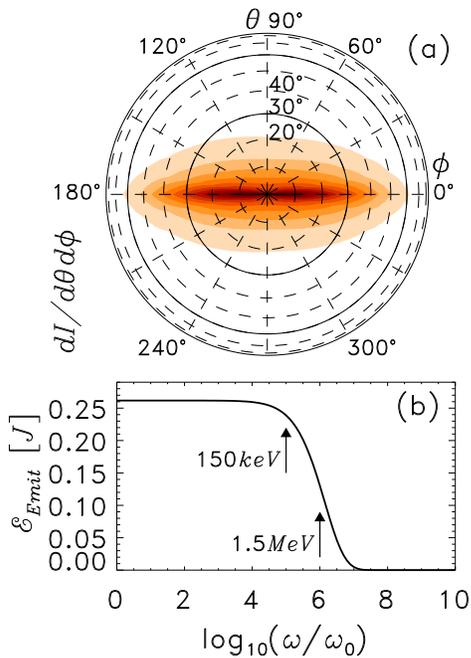}
\caption{
Results of 3D PIC simulation for a linearly polarized laser pulse with 
amplitude $a_0=70$ 
entering a soft plasma layer ($n_0=3n_{cr}$).
(a) Angular distribution of the backward scattered radiation with
photon energy above 150 keV 
and (b) total emitted energy as a function of the cut-off frequency.}  
\label{fig_4}
\end{figure}

In this simulation, which corresponds to the intensity
$10^{22}$W/cm$^2$ for $\lambda=0.8\mu m$, 
by the instant  $t=20\lambda/c$ the laser pulse loses $\sim 27$\% of its
energy, converting
$\sim 0.9$\% of the incident radiation (or $\sim 3.2$\% of the lost energy) 
to the backward scattered high-frequency radiation. 
The angular distribution of the radiation exceeding 150keV is shown in 
Fig.2(a). 
The emitted radiation has a wider angle along the direction of
a transverse electric field. 
The total energy of the pulse
equals $\sim$30 J, 
and emitted backward
high-frequency radiation accounts to $\sim$0.26 J,
with 0.24 J of photon energy above 150 keV (see Fig. 2(b)). 
The total radiated energy may be close to the 
particle energy for some electrons (see \cite{Zhidkov}). 
Specifically, this relates to 
a counter-streaming flow of electrons, 
with momentum up to $150mc$
propagating in the region of the laser pulse, as we observe in the
simulation. 
These fast electrons are generated in the charge separation field from
the flow of cold electrons; due to the loss of their energy in the
laser field, combined with an action of the charge separation field of 
opposite sign, they reverse their motion. 
However,  only a minor fraction of electrons counter-propagate with high 
energy, and of these,  only a fraction moves in the region of the strong laser 
field, where they can radiate. By this account the overall conversion 
efficiency is diminished and does not exceed the order of a few
percent. 

We have also applied the described model
within a particle-in-cell 
code 
to simulate 
processes pertinent to fast ignition at laser 
intensities $W\ge 10^{22}\ {\rm W/cm}^2$ \cite{inpress}. The processes in
higher fields, such that the corrected QED probabilities should be used to
simulate the emission, are to be considered in a forthcoming
publication.  
We also plan to use high contrast pulses from the Hercules laser to
drive high-density targets with intensities $>10^{22}$W/cm$^2$. We
hope that such a study may improve understanding of ultra-intense 
laser-plasma interactions  and may result in short X- or
$\gamma$-burst production.    

This work was supported by: the NSF (grant 0114336) and the ARO (grant DAAD19-03-1-0316).

\end{document}